\documentclass[aps,prd,twocolumn,superscriptaddress,nofootinbib,showpacs,floatfix]{revtex4-1}

\usepackage[pdftex]{graphicx}
\usepackage{epsfig}
\usepackage{amsmath}
\usepackage{amssymb}
\usepackage{dcolumn}
\usepackage{multirow}

\newcommand{\beq}{\begin{equation}}
\newcommand{\eeq}{\end{equation}}
\newcommand{\bear}{\begin{eqnarray}}
\newcommand{\eear}{\end{eqnarray}} \newcommand{\ba}{\begin{array}}
\newcommand{\ea}{\end{array}}

\begin{document}

\title{Constraints on Enhanced Dark Matter Annihilation from IceCube Results} 
\author{Ivone F.~M. Albuquerque} 
\email{ifreire@if.usp.br}
\author{Leandro J. Beraldo e Silva} 
\email{lberaldo@if.usp.br}
\affiliation{Instituto de F\'{i}sica, Universidade de S\~ao Paulo, S\~ao Paulo, Brazil}
\author{Carlos P\'erez~de los Heros}
\affiliation{Department of Physics and Astronomy,  Uppsala University,  Uppsala, Sweden}
\email{cph@physics.uu.se}

\date{\today}

\begin{abstract}
Excesses on positron and electron fluxes measured by ATIC, and the PAMELA and Fermi--LAT 
telescopes can be explained by dark matter annihilation in our Galaxy. However, this
requires large boosts on the dark matter annihilation rate. There are many possible 
enhancement mechanisms, such as the Sommerfeld effect or the existence of dark matter 
clumps in our halo. If enhancements on the dark matter annihilation cross section are taking place, 
the dark matter annihilation in the core of the Earth should also be enhanced. Here we use recent 
results from the IceCube 40--string configuration to probe generic enhancement scenarios. We present 
results as a function of the dark matter--proton interaction cross section, $\sigma_{\chi p}$ weighted 
by the branching fraction into neutrinos, $f_{\nu\overline{\nu}}$, as a function of a generic boost 
factor, $B_F$, which parametrizes the expected enhancement of the annihilation rate. We find that dark 
matter models which require annihilation enhancements of $\mathcal{O}(100)$ or more
and that annihilate significantly into neutrinos are excluded as the explanation for 
these excesses. We also determine the boost range that can be probed by the full IceCube telescope. 
\end{abstract}


\maketitle

\section{\label{introduction} Introduction}

Several recent observations in our galaxy have found an excess 
on the electron and/or positron fluxes~\cite{atic,pamela,fermilat}. However the origin of these events 
is yet not clear. If they are produced accordingly to standard physics, pulsars~\cite{puls} might be their
sources as well as they could be secondary particles created by shock accelerated hadrons~\cite{blasi}. 
Another exciting possibility is that they can be a signature of new physics. It was shown that dark 
matter annihilation in the galaxy can describe the data~\cite{bmod}. However most of these models 
require an enhancement on its annihilation rate. An enhancement mechanism can be found from a 
variety of possible phenomena, such as the Sommerfeld effect, the existence of dark matter 
substructures in the galactic halo, or a combination of both~\cite{cholis2, lattanzi}. \par

 Constraints on a boost factor on the dark matter annihilation rate have been recently derived from the 
Fermi--LAT diffuse gamma--ray measurement~\cite{fermicq} and from their analysis of Milky Way satellite galaxies~\cite{fermibnds}. 
An independent and neat analysis has been performed by the authors of Ref~\cite{cmbbnds}, using the 
fact that dark matter annihilations at recombination time, redshift $\sim$ 1000, would have injected secondary particles 
that would have affected the recombination processes. The measured power spectrum of the cosmic microwave background (CMB) 
can thus be used to set limits on the strength of the dark matter self--annihilation cross section. 
 The IceCube collaboration has also performed an analysis searching for dark matter annihilations in the galactic halo by 
searching from an excess diffuse neutrino flux over the expected atmospheric flux~\cite{ic22halo}. 
The results can also be used to set limits on a boost in the annihilation cross section.\par

 These methods show some degree of complementarity since each one provides better sensitivity to a different range of the mass of 
the dark matter particles or to different annihilation channels. They also rest on different assumptions and approximations. Fermi and 
the CMB method are competitive in the low mass range (dark matter massed $\lesssim$ 1~TeV), while IceCube reaches 10 TeV, and can also 
probe annihilation directly into neutrinos. There is some halo-model dependency when setting limits on the annihilation cross section 
from the observations of satellite galaxies: the expected signal depends on the degree of cuspiness of the assumed halo. 
On the other hand, the CMB analysis does not depend  on the shape of dark matter 
haloes, since there are no gravitationally bound structures at the redshift considered, but it depends on assumptions about the 
fraction of the energy released by the annihilating dark matter particles and how it is absorbed by the surrounding medium. 
\par
 
 In this paper we follow the alternative approach proposed in Ref.~\cite{paddy}. The authors argue that an enhancement of the dark 
matter annihilation rate should also boost the neutrino flux from dark matter annihilations in the center of the Earth. A thermal 
annihilation cross section implies that dark matter capture and annihilation
has not yet reached equilibrium in the Earth. However it is not necessary that the annihilation 
cross section has to remain the same as during the freeze--out period. If the annihilation 
rate is somehow enhanced in the post freeze--out period,
the equilibrium might already have been achieved. As it will be shown in the next Section,
the annihilation rate reaches its maximum at the equilibrium, and then depends only in the
capture rate. A boost in the annihilation rate increases the flux of annihilation
products until it reaches its maximum. In this case the neutrino flux from the center of the 
Earth will be large enough to be detected by telescopes such as IceCube. In short, if
the dark matter annihilation rate is enhanced, the timescale for equilibrium diminishes and the flux 
of annihilation products is expected to be much larger than the one away from equilibrium.\par

We use the recently published IceCube results on a search for a diffuse flux of muon neutrinos~\cite{ic40diff} to set limits on 
a boost factor in the dark matter annihilation cross section. 
IceCube has measured neutrinos coming from near or below the horizon in the energy range 332~GeV and 
84~TeV using data taken with the 40--string detector configuration (IceCube--40). The analysis includes all events 
coming from near or below the horizon, and the result is compatible with 
the expected atmospheric neutrino flux (see also Ref.~\cite{ic40atm}). It is also generic enough
to allow comparison with our prediction of the flux of muon neutrinos produced in dark matter
annihilations in the center of the Earth. We determine this flux by simulating the annihilation 
of WIMP--type particles in the center of the Earth and propagating the neutrinos to the detector. 
A significant neutrino flux from dark matter annihilation should be seen above the expected atmospheric neutrino flux. 
If not, limits can be set on the model used to determine the dark matter signal. Our analysis shows that models which 
require very large boosts on the dark matter annihilation rate in order to explain the excess seen in the galactic 
positron and electron flux, and have an annihilation channel into neutrinos, are ruled out.  \par

In the next Section we describe dark matter capture and annihilation in the Earth. In Sections~\ref{sec:neutrinos} 
and~\ref{sec:constrains} we describe the signal simulation and calculate the expected number of events in 
the IceCube--40 detector. We then compare our results with the IceCube--40 published results,
showing the boost factor range that is excluded. Finally, in Section~\ref{sec:prediction} we estimate the sensitivity 
region for the full 86--string detector.

\section{\label{sec:annihilation} Dark matter annihilation in the Earth}

Dark matter interactions in the Earth will be dominated by spin--independent elastic
scattering, since the most abundant isotopes of the Earth core and mantle are spin 0 
nuclei. The time evolution of the number of dark matter particles 
will result from a balance between the capture ($\Gamma_C$) and annihilation rate
($\Gamma_A$):
\beq
 \dot{N} = \Gamma_C - 2 \Gamma_A . 
\label{eq:dmevo}
\eeq
The Earth dark matter capture rate is given by~\cite{kam,paddy}:

\beq
\begin{split}
\Gamma_{C } \;\simeq \; 9.6 \times 10^{11} \frac{\rho_{\chi}}{0.3 {\rm GeV/cm^3}} 
\left(\frac{ 270 {\rm km/s}}{v_\chi}\right)^3  \\
\times  \left(\frac{{\rm TeV}}{m_\chi}\right)^2
\left(\frac{\sigma_{\chi p}}{10^{-42} {\rm cm^2}}\right) {\rm s}^{-1},
\end{split}
\label{eq:cpt}
\eeq
here $\sigma_{\chi p}$ is the spin--independent dark matter cross section off protons, $v_\chi$ 
and $\rho_\chi$ are the dark matter velocity and density in the halo and $m_\chi$ is the 
dark matter particle mass. In deriving the above equation it has been assumed that for 
masses m$_\chi$ much larger than the nucleus mass, the dark matter interaction cross section off a 
proton is the same as off a neutron. In this case, the spin--independent cross section off a nucleus with 
mass number A is given by $\sigma_{\chi N} \sim A^4\,\sigma_{\chi p}
(1-2m_p/m_\chi)$, where $m_p$ is the proton mass.\par

The annihilation rate depends both on the relative velocity scaled cross section 
$\left <\sigma_A v\right >$ as well as on the dark matter distribution in the Earth \cite{seckel}. 
The latter can be given in terms of the parameter 
$C_A\,\equiv \, \left <\sigma_A v\right >/V_{eff}$, where $\Gamma_A = N^2 C_A/2$, and 
$V_{eff} = 5.7 \times 10^{22} (TeV/m_\chi)^{3/2}$~cm$^3$ represents the dark matter effective 
volume in the core of the Earth, assuming an isothermal distribution \cite{gould,kam}. 

The solution to the dark matter time evolution in the Earth (Equation~\ref{eq:dmevo}) is then,
\beq
\Gamma_A \;=\; \frac{\Gamma_{C }}{2} \tanh^2\left(\frac{t_\oplus}{\tau}\right)
\label{eq:rann}
\eeq
where $t_\oplus$ is the age of the solar system and $\tau = 1/\sqrt{\Gamma_c C_A}$ is the
timescale for equilibrium between capture and annihilation. An enhancement on the 
annihilation rate will only be effective if it can accelerate equilibrium within the Earth. 
When the equilibrium stage is reached, the annihilation is maximum and depends entirely
on the capture rate ($\Gamma_A \,=\, \Gamma_C/2$). Thermal relic dark matter candidates
typically have $\left <\sigma_A v\right >_r = 3 \times 10^{-26}$~cm$^3$s$^{-1}$, which makes
$C_{Ar} \,\simeq \, 5.3 \times 10^{-49} (m_\chi/TeV)^{3/2}$~s$^{-1}$ and for these values 
the Earth is today far from equilibrium. \par

In our analysis we consider scenarios where the annihilation cross section is enhanced 
by boosting the thermal relic annihilation rate $\Gamma_{Ar}$ by a generic factor $B_f$ 
which affects the non--equilibrium rate through $C_A \,=\, B_f C_{Ar}$. Such parametrization, though 
adequate to probe enhancements due to Sommerfeld effect or to new interaction mechanisms, cannot 
probe enhancements due to a possible dark matter halo substructure. In this latter case, a standard 
annihilation cross section could well account for any possible signal, which would be just due to the 
increased local dark matter density, and not due to any new feature of
the annihilation process itself.

\section{\label{sec:neutrinos} Neutrino Production}

The escape velocity at the Earth core is $v_e \approx 15$~km/s and dark matter moves very slowly
inside the Earth. The annihilation products will therefore be monochromatic, and produced with 
the same energy as the dark matter mass. Here we consider the annihilation into a muon neutrino pair   
(which we call primary neutrinos). ``Secondary'' neutrinos are also produced in the decay of other primary 
annihilation particles, such as $\tau$'s, t's, W's and b's. Since the explanation of the observed leptonic 
excess in terms of dark matter has also to account for the fact that no antiproton
excess was found by PAMELA~\cite{pamantip}, annihilation into leptons is preferred
when compared to hadrons. Secondary neutrinos from 
annihilation into charged lepton states, specifically on $\tau\bar{\tau}$ were
analyzed in Ref.~\cite{paddy}. The energies of these secondary neutrinos will be spread at relatively low values 
($\sim 50$~GeV) compared to the primary neutrino flux, and detection in neutrino telescopes is then disfavored 
(unreasonably large boost factors would be needed to bring such flux over the atmospheric neutrino background to 
a detectable level). We will therefore not take into account the secondary neutrino flux in our calculations, and 
concentrate in the easily detectable monochromatic flux from direct annihilations.\par

The primary neutrino flux, produced from dark matter annihilations in the Earth's center is given by:
\beq
\frac{d\phi_\nu}{dE_\nu\,dA\,dt\,d\Omega} \;=\; \frac{f_{\nu\overline{\nu}} \; \Gamma_A}{4 \; \pi \; R_\oplus^2}\frac{dN_\nu}{dE_\nu}
\label{eq:nuflx}
\eeq
where $f_{\nu\overline{\nu}}$ is the annihilation branching ratio into 
$\nu_\mu\overline{\nu}_\mu$, $R_\oplus$ is the radius of the Earth, and $dN_\nu/dE_\nu$ is the 
energy distribution of the neutrinos produced in the annihilations. We show our results for 
two generic cases, $m_\chi = 500$~GeV and 
$m_\chi = 1000$~GeV, which are representative of the models which fit the observed positron and 
electron excess, and later on for $m_\chi = 5000$~GeV when comparing our results 
to others. Since at these energies neutrinos practically do not lose energy in their way from 
the center of the Earth to the detector, the term ${dN_\nu}/{dE_\nu}$ will then be a delta function 
at the dark matter masses considered.\par

\begin{figure}[t]
\includegraphics[width=\linewidth,height=7.5cm]{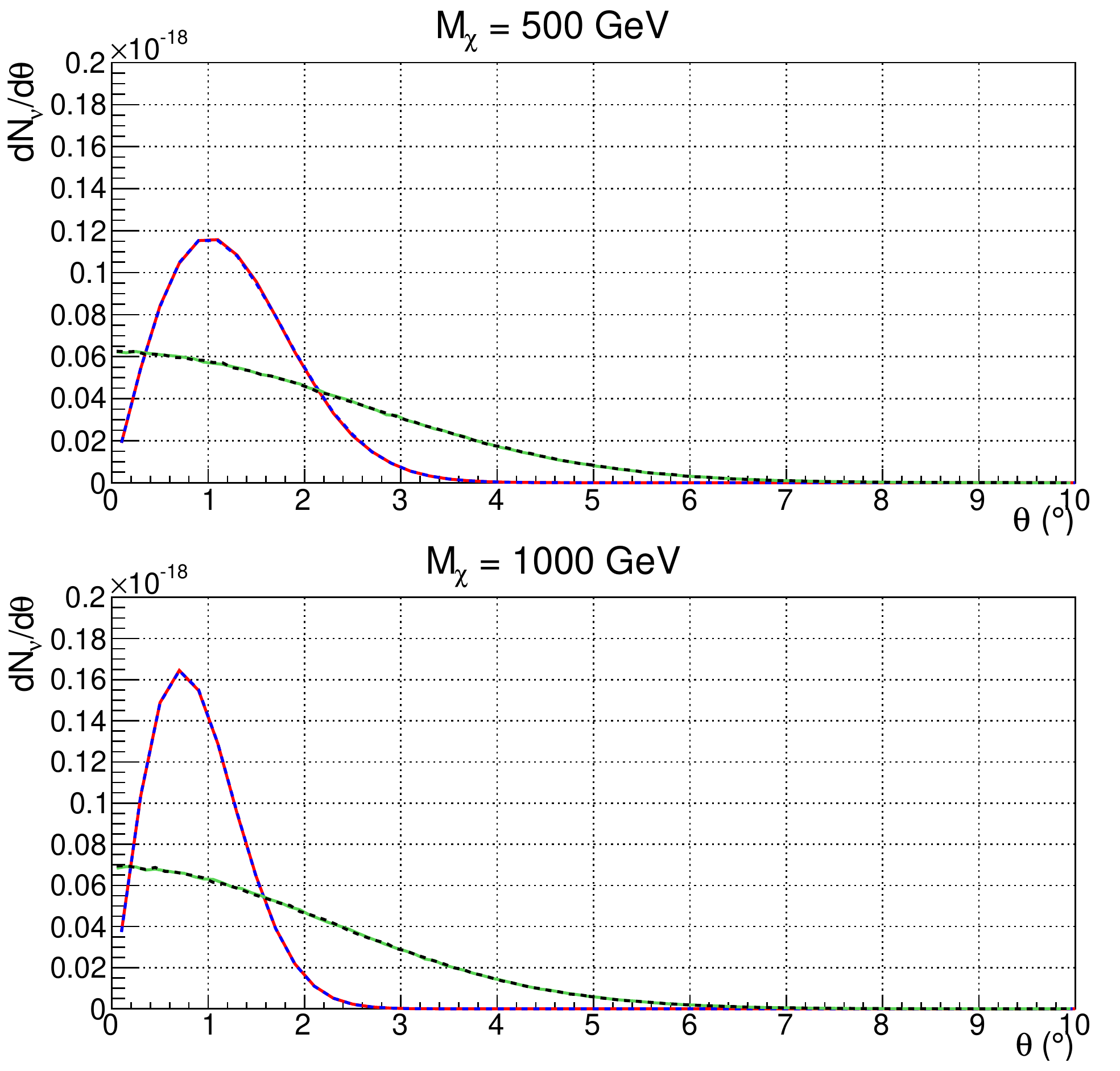}
\caption{ Angular distributions of the primary muon neutrino and antineutrino flux 
from dark matter annihilations at the center of the Earth with respect to the vertical
direction (pink lines). The green curves show the angular distribution smeared by the 
detector angular resolution, taken as 2$^\circ$. The bottom plot shows the distribution 
for 1~TeV dark matter, while the plot on the top for 500~GeV dark matter. Full lines are
for neutrinos and dashed are for antineutrinos.}
\label{fig:nutht}
\end{figure}

\section{\label{sec:constrains} Constraints on the annihilation boost factor from IceCube--40}

In order to predict the muon flux from dark matter annihilation in the Earth at the IceCube 
detector in the South Pole, we use the publicly available WimpSim code~\cite{wimpsim}.
We simulate a monochromatic muon neutrino beam at the center of the Earth, 
with energy equal to the dark matter mass, by selecting the muon neutrino channel. WimpSim simulates 
the propagation of these neutrinos, including energy losses and charged and neutral current 
interactions as well as oscillations through the Earth. The muon neutrino flux at the 
detector is given as an output. Although we have chosen to simulate the flux  
specifically for the location of the IceCube detector, the detector site is not relevant  
in this case, and the results and sensitivities presented in the next sections can easily be 
interpreted for any neutrino telescope of similar size.\par

 The number of muons from dark matter annihilation from a given angular region $\Omega$ and during 
an exposure $t_{\rm exp}$ in IceCube--40, is then:
\beq
N_\mu \;=\; \int \frac{d\phi_\nu}{dE_\nu\,dA\,dt\,d\Omega} \; dE_\nu \; t_{\rm exp} \; A_{\rm eff}\; \Omega.
\label{eq:nevts}
\eeq

We proceed then by convoluting the muon neutrino flux with the IceCube--40 effective area published 
in Ref.~\cite{ic40diff}. The effective area accounts for the detector efficiency including the 
neutrino--nucleon interaction probability, the muon energy loss from its production point to the 
detector, and the detector trigger and analysis efficiency. As the  neutrino beam is monochromatic, 
we use the corresponding value of $A_{\rm eff}$ for each dark matter mass considered. 
The muon--neutrino angular distribution is the main parameter for background reduction in our 
analysis. Figure~\ref{fig:nutht} shows the angular distribution from neutrinos from dark matter 
annihilation in the center of the Earth. It is collimated in a angle of less than approximately 
$3^\circ$ around the vertical direction, $\theta = 180^\circ$. This figure also shows the smearing effect 
due to the angular resolution of the detector, taken here conservatively to be 2$^\circ$ for up--going 
vertical events. We use therefore the effective area for the zenith range 
$150^\circ < \theta < 180^\circ$ from Ref.~\cite{ic40diff}. Even though such angular range is much wider 
than that from the expected signal, the effective area does not vary considerably with zenith angle 
below about 10 TeV, where Earth absorption effects start to be important. \par

\begin{figure}[t]
\includegraphics[width=\linewidth,height=7.5cm]{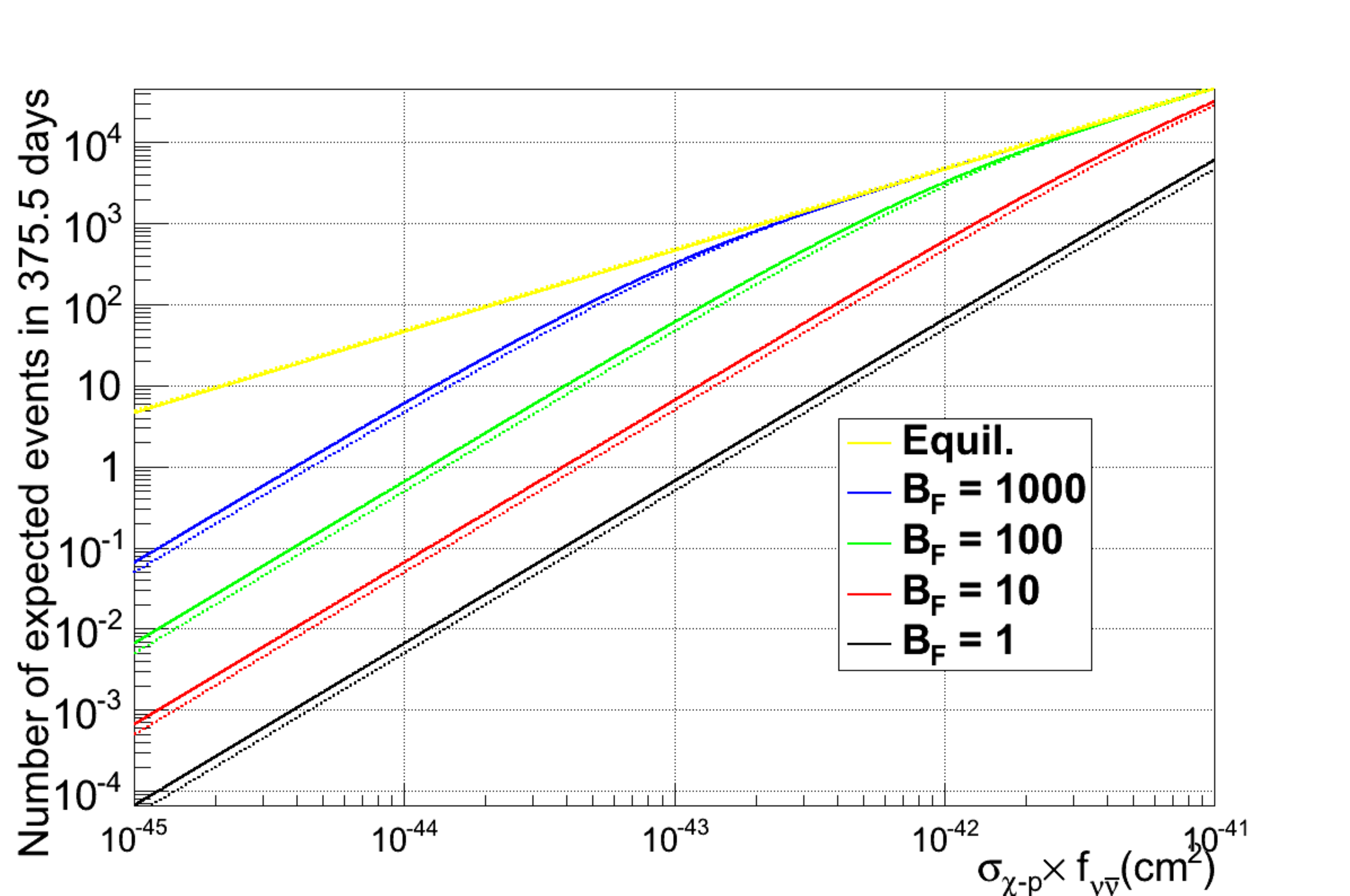}
\caption{Predicted number of $\nu_\mu + \overline{\nu}_\mu$ from dark matter 
annihilation in the Earth at IceCube--40. Results are for 
375.5 days of exposure for different boost factor values (color coded as labeled). Dashed 
lines are for 1 TeV and solid lines for 500 GeV dark matter.} 
\label{fig:nmu}
\end{figure}

 For our analysis, we choose an angular window from the vertical direction that contains 90\% of the 
expected signal ($4.1^\circ$($3.7^\circ$) for 500(1000)~GeV dark matter respectively, where the angular
experimental resolution has been taken into account), and use the calculated number of events 
with Eq.~\ref{eq:nevts} as signal. 
We have not used any energy information in this analysis, just using the total number of events predicted 
and detected to derive our exclusion regions. We will however present a study on the effect of the detector energy resolution in the sensitivity study 
for the complete IceCube detector in the next section. Figure~\ref{fig:nmu} shows the predicted number of 
$\nu_\mu + \overline{\nu}_\mu$ in IceCube--40, with $t_{\rm exp} = 375.5$ days as a function of the dark 
matter nucleon cross section $\sigma_{\chi p}$ scaled by the branching ratio for different boosts 
factors. As can be seen, larger enhancements on the annihilation rate result in a shorter time for 
equilibrium among capture and annihilation.\par

In order to study the model rejection power of our analysis in a quantitative way, we 
determine the statistical significance of the predicted dark matter signal, $S/\sqrt{B}$,  
as a function of boost factor, where $S$ is the number of signal events predicted in the given 
angular cone. As background, $B$,  we use the measured number of events in IceCube--40 
in the same angular regions. \par

Two recent IceCube publications, a measurement of the atmospheric neutrino 
flux~\cite{ic40atm} and a search for a diffuse $E^{-2}$ flux of cosmic origin~\cite{ic40diff}, 
give results that are consistent  with the expected atmospheric neutrino flux. 
We use the publicly available data from the diffuse analysis 
which is available at Ref.~\cite{icedata}. This analysis selects events coming from near or 
below the horizon and their background is composed by atmospheric muons arriving 
from above the detector and misreconstructed as an event coming from below. The rejection of 
these events is described in Ref.~\cite{ic40diff} and the background contamination in the data 
sample is estimated to be less than 1\%. The final data sample contains 13K events and it 
is given as a function of the zenith angle.  This allows us to select events which come in the 
same direction as expected signal dark matter annihilations. The data is then reduced 
to 14 (9) events when considering the angular regions expected for 500 (1000)~GeV dark 
matter neutrinos. We compare this number to the predicted number of events from 
our signal choices and for different boost factors. \par

\begin{figure*}[t]
\includegraphics[width=\linewidth]{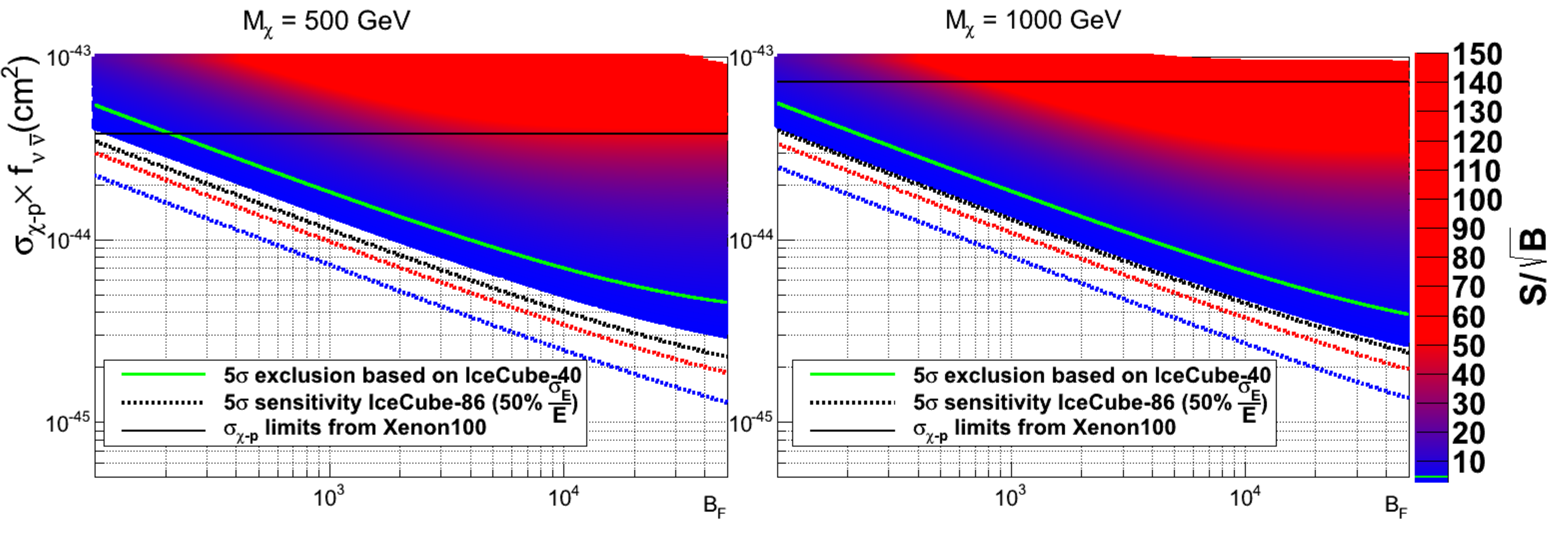}
\caption{Boost factor versus dark matter nucleon interaction cross section $\sigma_{\chi p}$ scaled by 
$\nu\overline{\nu}$ branching ratio ($f_{\nu\overline{\nu}}$). The light green full line indicates the 
exclusion at $5 \sigma$ level based on our analysis of IceCube--40 results. 
Above this line the exclusion is more strict accordingly to the color coded bar at the side. 
The color coded bar represents the statistical significance level ($S/\sqrt{B}$). 
Both the bar as well as the contour in this figure starts at $3 \sigma$. 
The black (red, blue) dashed line is the $5 \sigma$ sensitivity for the IceCube--86 telescope
with 50 (30, 10)\% energy resolution. 
As a reference, the black full lines represent the 90\% CL constraints on $\sigma_{\chi p}$ from direct 
detection~\cite{xe100} (independent of the boost factor and of the annihilation branching ratio 
$f_{\nu\overline{\nu}}$). The plot to the left is for 500 GeV dark matter and the one to the right for 
1000 GeV.
}
\label{fig:cl} 
\end{figure*}

The result is presented in Fig.~\ref{fig:cl}, 
which shows the boost factor versus $\sigma_{\chi p}$ scaled by the $\nu\overline{\nu}$ branching 
ratio region which was probed in this analysis. The plot to the left is for 500 GeV dark matter 
and the one to the right is for 1 TeV. 
The light green full line shows a $5 \sigma$ exclusion and above this line the exclusion 
is more strict according to the color coded bar on the right side. This bar indicates the 
level of statistical significance. The best limit on $\sigma_{\chi p}$ comes from the Xenon100 
collaboration~\cite{xe100}. At 90\% CL 500 GeV dark matter models are constrained above $\sigma_{\chi p} \approx 4 \times 
10^{-44}$~cm$^2$ and 1 TeV dark matter above $\sigma_{\chi p} \approx 8 \times 10^{-44}$~cm$^2$. We draw this 
limit on Fig.~\ref{fig:cl} just as a reference, since direct detection results do not depend on $f_{\nu\overline{\nu}}$ 
neither on any boost factor. For such limit, boost factors above 215 and 58 respectively are excluded at a $5\sigma$ level.
This exclusion requires, however, that $f_{\nu\overline{\nu}} =1$, ie., annihilation 
exclusively into neutrinos, which may be difficult to justify phenomenologically.  But it should be noted  that 
the excluded region in Fig.~\ref{fig:cl}  does not require necessarily a 100\% annihilation into neutrinos. 
One can lower the branching factor by increasing the required boost factor limit.
As one moves down along the green 5$\sigma$ limit line, larger boost factors are disfavored, as the quantity  
$\sigma_{\chi p} \times f_{\nu\overline{\nu}}$ becomes lower. Lower values 
of  $\sigma_{\chi p} \times f_{\nu\overline{\nu}}$ can be achieved either by models with relatively high cross sections
but a lower branching ratio to neutrinos, or by lower cross sections but a higher annihilation probability 
to neutrinos. In either case, the expected signal in IceCube would decrease and a higher boost factor 
would be needed to bring it to the current sensitivity of the detector.
Higher boost factors to the right of the green line would then be disfavored at a higher significance.

The degeneracy of the product $\sigma_{\chi p} \times f_{\nu\overline{\nu}}$ can be 
broken when considering specific dark matter 
models with known branching ratio to neutrinos and cross section with protons.  
Fig.~\ref{fig:cl} can then be used to determine the  minimum boost factor which is disfavored at a 5$\sigma$ level.

\section{\label{sec:prediction} Prediction for full IceCube detector}

The full IceCube detector with 86 strings (IceCube--86) is now completed and taking data 
since May 2011.  In order to estimate the full detector sensitivity, we can use the fact 
that the atmospheric neutrino flux measured by IceCube--40~\cite{ic40diff,ic40atm} is 
consistent with model expectations as, for example, the one proposed by Ref.~\cite{honda}. 
We will also go one step further than in the calculations of the previous Section, 
and assume that IceCube will be able to estimate the neutrino energy with a given 
resolution $\sigma_E$. We assume three benchmark energy resolutions, $\sigma_E=0.1E, 
0.3E$ and $0.5E$ which go from the very optimistic to the more conservative situation 
in a neutrino telescope. The 40--string configuration has its energy resolution  
between 50\% and 80\%. We assume here
that the complete detector can reach a better resolution in any case.
We note that energy estimation in neutrino telescopes is 
a difficult task since muon tracks above a few hunderd GeV will cross the detector volume,  
and the neutrino energy can only be estimated through model--dependent deconvolution 
methods.  \par

 An effective area at trigger level for the complete IceCube--86 detector has been published 
in Ref.~\cite{ic86}. However the effective area at final analysis level can differ significantly 
from trigger level, since data quality cuts are applied to the data sample to reduce background, 
but also inevitably reducing signal efficiency. In practice, one can just rescale the 
IceCube--40 effective area by 2.15 (which factors 40 to 86 strings), since at the energies 
considered in our analysis, and for vertical events, each string is practically an 
independent detector and therefore, in a first approximation, the effective area scales 
proportionally to just the number of strings. We expect, though, dedicated analysis 
with IceCube--86 to be more efficient at lower energies than the IceCube--40 
analysis we have used in the previous Section and, since we will be integrating 
 A$_{\rm eff}\times {\rm flux}$ in a range of energies, we need to be careful 
with the behavior of the effective area with energy. 
 We have therefore normalized the IceCube--86 effective area from Ref.~\cite{ic86} with 
the rescaled IceCube--40 area at 10 TeV, where the IceCube--40 analysis is optimal. 
The shape of the IceCube--86 effective area at lower energies takes automatically into account 
the improved capabilities of the full detector at energies below 1~TeV. \par

We have then all the ingredients we need to estimate the sensitivity of the IceCube--86 
detector. We calculate the number of atmospheric neutrino events  from 
within a vertical cone of 4.1$^\circ$ and 3.7$^\circ$ aperture by convoluting the 
detector effective area with the parametrization of the atmospheric neutrino flux 
taken from Ref.~\cite{honda}. Instead of choosing a delta function for the spectrum as 
in the previous case, we introduce a smearing in the energy according to the assumed energy 
resolution. In practice, this translates into that we perform the A$_{\rm eff}\times {\rm flux}$ 
integral between m$_{\chi}-\sigma_E$ and m$_{\chi}+\sigma_E$. We assume that the measurement of 
IceCube will be compatible with the calculated number of background events and 
calculate the sensitivity to an excess neutrino flux from the center of the 
Earth under this assumption. The result of this exercise is shown in  Fig.~\ref{fig:cl}. 
The dashed black (red,blue) lines in both plots represent $5 \sigma$ sensitivity that can be achieved 
in one year  live time of IceCube--86 ($t_{\rm exp86} = 365$ d) assuming a 50 (30,10)\%
energy resolution. 

\section{\label{sec:conclusions} Conclusions}

If dark matter annihilation in our galaxy is responsible for the excesses 
seen by ATIC~\cite{atic}, PAMELA~\cite{pamela} and Fermi--LAT~\cite{fermilat}, 
an enhancement on the dark matter annihilation rate in the Earth should also be foreseen. 
This enhancement might bring the equilibrium among the dark matter capture and annihilation 
rates to occur much earlier than the timescale expected from a  purely thermal annihilation 
rate. In this case the neutrino flux from dark matter captured in the Earth can be quite large. 
We present results as a function of the  dark matter--proton interaction cross section, 
$\sigma_{\chi p}$ weighted by the branching fraction into neutrinos, $f_{\nu\overline{\nu}}$, as a function 
of a generic boost factor, $B_F$, which parametrizes the expected enhancement of the annihilation rate.  
In this sense, it is important to note that our results do not depend on the details of the mechanism 
which enhances the annihilation cross section. 
We have used two benchmark models, a 500~GeV and 1~TeV generic WIMP annihilating in the center of 
the Earth to scan the  ($\sigma_{\chi p} \times f_{\nu\overline{\nu}}$, $B_F$) parameter space and 
set constrains on this 2-dimensional space using current IceCube results. Our calculations assume that 
the dark matter velocity distribution is Gaussian as well as that dark matter collected in the Earth's 
core follows an isothermal distribution. \par

In order to explain the positron and electron excesses that are seen, models also have to cope with the 
fact that the antiproton spectrum measured by PAMELA~\cite{pamantip} is in full agreement with the expectation 
from secondary production of antiprotons from propagation of cosmic rays in the galaxy. This rules out as an 
explanation of the excess many dark matter models with preferred annihilation into heavy products, producing 
many antiprotons. Therefore leptophilic models, which propose dark matter annihilation exclusively into leptons, 
are favored as an explanation to the excesses found. 
We have shown that, when 500 (1000)~GeV dark matter annihilates into a large 
fraction of neutrinos, annihilation boost factors of the $\mathcal{O}(100)$ and above are already 
constrained by our analysis at a 5$\sigma$ level, or higher, depending on the interaction 
cross section assumed. Thus, leptophilic models~\cite{leptop} which favor primary 
neutrino production are constrained by our results. \par

\begin{figure}[t]
\vspace*{-1.5cm}
\hspace*{-.5cm}
\includegraphics[width=4.15in,height=5.25in]{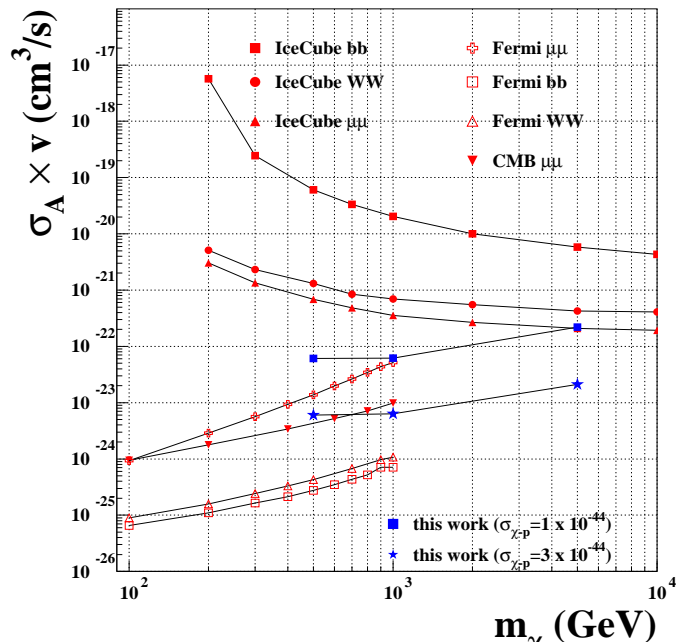}
\vspace*{-2.5cm}
\caption{Bounds on the dark matter annihilation cross section $\sigma_{A}$ versus
mass $m_\chi$ by
Fermi~\cite{fermibnds}, IceCube~\cite{icebnds}, analysis over CMB data~\cite{cmbbnds}
and this work, where two different $\sigma_{\chi p}$ values of 
$3 \times 10^{-44}$~cm$^2$ (blue star) and $1 \times 10^{-44}$~cm$^2$ (blue solid square) are assumed. Here we also plot
our results assuming a 5 TeV dark matter mass. Our limits correspond to a $3 \sigma$
significance level, while the others are at 90\% CL.} 
\label{fig:cmp} 
\end{figure}

In order to compare this analysis to others, we show in 
Figure~\ref{fig:cmp} limits on
$\left< \sigma_A \times v \right>$ as a function of $m_\chi$ from
Fermi~\cite{fermibnds}, CMB~\cite{cmbbnds} and IceCube~\cite{icebnds}.
Our bounds as a function of $\left< \sigma_A \times v \right>$, can be determined
from equation~\ref{eq:rann}. It can also easily be visualized in 
Figure~\ref{fig:cl}, where a $3 \sigma$ significance is shown as 
the edge of the shaded area: the limit on the annihilation cross section is determined
from each boost factor value on this curve. In Figure~\ref{fig:cmp} these are shown for two choices of 
$\sigma_{\chi p}$, 
$3 \times 10^{-44}$~cm$^2$ (blue stars) and $1 \times 10^{-44}$~cm$^2$ (blue solid 
squares), which are below the current Xenon limit~\cite{xe100}. 

Our results and IceCube's, are the only ones to probe 
annihilation into neutrinos. 
Other limits coming from searches with gammas from
satellite galaxies by Fermi~\cite{fermibnds} or from the analysis based on the 
CMB~\cite{cmbbnds} imprint at high redshifts,
are complementary, since not only
they probe different annihilation channels, but also rest on different 
underlying assumptions. \par

 We also investigated the reach of the completed IceCube 86-string detector, and we present 
results of its sensitivity in the  ($\sigma_{\chi p} \times f_{\nu\overline{\nu}}$, $B_F$) parameter space. We have 
estimated how using neutrino energy information could improve the analysis, and in Fig.~\ref{fig:cl} we show the 
expected sensitivity for three different assumptions of the detector energy resolution.

\acknowledgments
We thank P.~Fox and P.~Gouffon for fruitful discussions. IA was partially funded by 
the Brazilian National Counsel for Scientific Research (CNPq), and L.J.B.S. was partially 
funded by the State of S\~{a}o Paulo Research Foundation (FAPESP) and by CNPq.


\begin{thebibliography}{99}
\bibitem{atic} J.~Chang {\it et al,},
Nature {\bf 456} (2008) 362.
\bibitem{pamela}  O.~Adriani {\it et al,}, PAMELA Collaboration, 
[arXiv:1103.2880];
  Astropart.\ Phys.\  {\bf 34}, 1-11 (2010).
  [arXiv:1001.3522]; O.~Adriani {\it et al,}, PAMELA Collaboration,  
  Phys.\ Rev.\ Lett.\  {\bf 102}, 051101 (2009).
  [arXiv:0810.4994].
\bibitem{fermilat}A.~A.~Abdo {\it et al,},  Fermi--LAT Collaboration, 
  Phys. Rev. Lett. {\bf 102}, 1811101 (2009).
\bibitem{puls}  S.~Profumo,
Central Eur.\ J.\ Phys.\  {\bf 10}, 1 (2011)
  [arXiv:0812.4457 [astro-ph]].
\bibitem{blasi} P.~Blasi,
  Phys.\ Rev.\ Lett.\  {\bf 103}, 051104 (2009).
  [arXiv:0903.2794].
\bibitem{bmod} M.~Cirelli and A.~Strumia,
  PoS {\bf IDM2008}, 089 (2008).
  [arXiv:0808.3867];
I.~Cholis, L.~Goodenough, D.~Hooper, M.~Simet and N.~Weiner,
  Phys.\ Rev.\  D{\bf 80}, 123511 (2009).
  [arXiv:0809.1683];
  I.~Cholis, L.~Goodenough, D.~Hooper, M.~Simet and N.~Weiner,
  Phys.\ Rev.\   D{\bf80}, 123511 (2009).
  [arXiv:0809.1683]. 
D.~P.~Finkbeiner, L.~Goodenough, T.~R.~Slatyer, M.~Vogelsberger and N.~Weiner,
  JCAP {\bf 1105}, 002 (2011).
  [arXiv:1011.3082].
\bibitem{cholis2} I.~Cholis and L.~Goodenough,
  JCAP {\bf 1009}, 010 (2010).
\bibitem{lattanzi} M.~Lattanzi and J. Silk, 
  Phys. Rev. D{\bf 79}, 083523 (2009).
\bibitem{fermicq}  A.~A.~Abdo {\it et al,},  Fermi--LAT Collaboration,
  JCAP {\bf 1004}, 014 (2010), [arXiv:1002.4415];
\bibitem{cmbbnds} S.~Galli, F.~Iocco, G.~Bertone and A.~Melchiorri,
  Phys.\ Rev.\ D {\bf 84}, 027302 (2011).
  [arXiv:1106.1528 [astro-ph.CO]];
  G. H\"utsi, J. Chluba, A. Hektor and M. Raidal,
  Astron. \& Astrophys. {\bf 535}, A26 (2011).
\bibitem{ic22halo} R. Abbasi  {\it et al,} IceCube Collaboration, 
  Phys.\ Rev.\  D{\bf 84}, 022004 (2011).
\bibitem{paddy} C.~Delaunay, P.~J.~Fox and G.~Perez,
  JHEP {\bf 0905}, 099 (2009).
  [arXiv:0812.3331].
\bibitem{ic40diff} R.~Abbasi {\it et al,},  IceCube Collaboration,
   Phys. Rev. D{\bf 84}, 082001 (2011).
    [arXiv:1104.5187].
\bibitem{ic40atm} R.~Abbasi {\it et al,} IceCube Collaboration, 
  Phys.\ Rev.\  D{\bf 83}, 012001 (2011).
  [arXiv:1010.3980].
\bibitem{kam} G.~Jungman, M.~Kamionkowski and K.~Griest,
  Phys.\ Rept.\  {\bf 267}, 195-373 (1996).
  [hep-ph/9506380].
\bibitem{seckel} K.~Griest and D.~Seckel,
  Nucl.\ Phys.\  B{\bf 283}, 681 (1987).
\bibitem{gould}   A.~Gould,
  Astrophys.\ J.\  {\bf 321}, 571 (1987).
\bibitem{pamantip} O.~Adriani {\it et al,}, PAMELA Collaboration, 
  Phys.\ Rev.\ Lett.\  {\bf 105}, 121101 (2010).
  [arXiv:1007.0821].
\bibitem{wimpsim} M.~Blennow, J.~Edsj\"o and T.~Ohlsson,
  JCAP {\bf 0801}, 021 (2008).
  [arXiv:0709.3898].
\bibitem{icedata} {\url{http://www.icecube.wisc.edu/science/data/}}.
\bibitem{honda} M.~Honda {\it et al,} Phys.\ Rev.\  D{\bf 75}, 043006 (2007).
  [astro-ph/0611418].
\bibitem{feld} G.~Feldman and R.~D.~Cousins,
  Phys.\  Rev.\ D{\bf 57}, 3873 (1998).
\bibitem{xe100} E.~Aprile {\it et al,}, XENON100 Collaboration, 
  Phys. Rev. Lett. {\bfseries 107}, 131302, (2011).
  [arXiv:1104.2549].
\bibitem{cdms} Z.~Ahmed {\it et al,}, CDMS--II Collaboration, 
  Science {\bf 327}, 1619-1621 (2010).
  [arXiv:0912.3592].
\bibitem{ic86} C.~Wiebusch, for the IceCube Collaboration,
  Procc. of the 31st ICRC, Lodz, Poland, July 2009
  [arXiv:0907.2263].
\bibitem{leptop} P.~J.~Fox and E.~Poppitz,
  Phys.\ Rev.\  D{\bf 79}, 083528 (2009).
  [arXiv:0811.0399].
\bibitem{Abbasi:2011eq} 
  R.~Abbasi {\it et al,}  [IceCube Collaboration],
  Phys.\ Rev.\ D {\bf 84}, 022004 (2011)
  [arXiv:1101.3349 [astro-ph.HE]].
\bibitem{fermibnds} A.~Ackermann {\it et al,},  Fermi--LAT Collaboration, 
 Phys. Rev. Lett. {\bf 107}, 241302 (2011).
\bibitem{icebnds} A. Abbasi {\it et al,},  IceCube Collaboration,
Phys. Rev. D{\bf 84}, 022004, (2011);  see also [arXiv:1111.2738]
\end{thebibliography}
\end{document}